\documentclass[conference]{IEEEtran}
\IEEEoverridecommandlockouts
\usepackage{cite}
\usepackage{amsmath,amssymb,amsfonts}
\usepackage{algorithmic}
\usepackage{graphicx}
\usepackage{textcomp}
\usepackage{makecell}
\usepackage{listings}
\usepackage{hhline}

\usepackage{xcolor}
\def\BibTeX{{\rm B\kern-.05em{\sc i\kern-.025em b}\kern-.08em
    T\kern-.1667em\lower.7ex\hbox{E}\kern-.125emX}}

\begin{document}

\title{Exploring the acceleration of Nekbone on reconfigurable architectures}

\author{\IEEEauthorblockN{Nick Brown}
\IEEEauthorblockA{\textit{EPCC at the University of Edinburgh} \\
The Bayes Centre, 47 Potterrow, Edinburgh \\
n.brown@epcc.ed.ac.uk}
}

\maketitle

\begin{abstract}
Hardware technological advances are struggling to match scientific ambition, and a key question is how we can use the transistors that we already have more effectively. This is especially true for HPC, where the tendency is often to throw computation at a problem whereas codes themselves are commonly bound, at-least to some extent, by other factors. By redesigning an algorithm and moving from a Von Neumann to dataflow style, then potentially there is more opportunity to address these bottlenecks on reconfigurable architectures, compared to more general-purpose architectures.

In this paper we explore the porting of Nekbone’s AX kernel, a widely popular HPC mini-app, to FPGAs using High Level Synthesis via Vitis. Whilst computation is an important part of this code, it is also memory bound on CPUs, and a key question is whether one can ameliorate this by leveraging FPGAs. We first explore optimisation strategies for obtaining good performance, with over a 4000 times runtime difference between the first and final version of our kernel on FPGAs. Subsequently, performance and power efficiency of our approach on an Alveo U280 are compared against a 24 core Xeon Platinum CPU and NVIDIA V100 GPU, with the FPGA outperforming the CPU by around four times, achieving almost three quarters the GPU performance, and significantly more power efficient than both. The result of this work is a comparison and set of techniques that both apply to Nekbone on FPGAs specifically and are also of interest more widely in accelerating HPC codes on reconfigurable architectures.
\end{abstract}


\section{Introduction}
Scientists are placing unprecedented demands upon supercomputers, requiring the ability to address larger problems at reduced time to solution. Whilst for a number of years there has been a trend to adopt technologies that provide increased raw computational power, for instance GPUs or CPUs with more advanced vector units such as SVE, many HPC codes are not fully bound by computation. Therefore, a key question is whether there is benefit in leveraging an architecture that enables us to directly address some of the other performance bottlenecks in High Performance Computing (HPC) codes. One potential advantage of reconfigurable architectures is that, by specialising the electronics to a specific application, then key details, such as how memory is accessed, can be tuned on a code by code basis, rather than relying upon architectural decisions which are more general purpose.

Field Programmable Gate Arrays (FPGAs) are a form of reconfigurable architecture which provide a large number of configurable logic blocks sitting within a sea of configurable interconnect. With the addition of other facets on the chip, such as fast memory which includes block RAM (BRAM) and UltraRAM (URAM), Digital Signal Processing (DSP) slices, and high bandwidth connections off-chip, FPGAs are hugely versatile. The adoption of this technology in scientific computing has, until now, been rather limited, but the significant investment made by vendors in the recent years around the software ecosystem for programming FPGAs, along with the availability of more capable hardware, has the potential to drive increased ubiquity.

Whilst even the most powerful FPGAs struggle to compete with latest CPUs or GPUs when it comes to raw floating point capability, for codes which are not fully computationally bound then potentially other factors are more crucial in dictating performance. A key question is whether, by reformulating algorithms in dataflow form, and running on reconfigurable architectures, one can offset other bottlenecks such as memory overheads. By doing so one is then keeping the FPGA's DSP slices continually busy performing floating point operations, and without stalling this makes more effective use of the transistors that we already have. Put simply, our general hypothesis is that the ability to specialise reconfigurable architectures at the electronics level with dataflow algorithms will enable the hardware to be kept busy in situations where other more general purpose Von Neumann based architectures would stall. Furthermore, FPGAs promise significant power efficiency benefits, and a secondary question is whether such architectures can provide improved performance per Watt over technologies more commonly found in HPC machines.

In November 2019, Xilinx released their Vitis platform which aims to make the programming of FPGAs more a question of software development rather than hardware design. In this paper we use Vitis to accelerate Nekbone, a popular HPC mini-app, which captures the basic structure of the Nek5000 application, on a Xilinx Alveo U280 FPGA. Specifically, we focus on the AX kernel, which accounts for over 75\% of the runtime, and this paper is structured as follows; in Section \ref{sec:background} we explore the background to this work, describing the FPGA technologies we are using, the Nekbone mini-app in more detail, and the hardware itself. Section \ref{sec:kernel_optimisation} then describes optimisation at the single FPGA kernel level, transforming the algorithm from a Von Neumann based CPU version to the dataflow style. We then focus on scaling up the number of FPGA kernels in Section \ref{sec:multiple_kernel}, comparing runs on a Xilinx Alveo U280 against the performance and power efficiency of an Intel Xeon Platinum CPU and a NVIDIA Tesla V100 GPU, before concluding and discussing further work in Section \ref{sec:conclusions}.


\section{Background}
\label{sec:background}

In recent years there has been much development in programming tools for FPGAs, and the use of higher level programming abstractions such as High Level Synthesis (HLS) is amongst the most prevalent of these. When using HLS, a kernel is written in C, C++ or System C and then translated by the tooling into the underlying Hardware Description Language (HDL). Freeing programmers from having to work at the HDL level directly increases productivity and opens the technology up to more of the software community. Within their high level code, programmers are able to direct the tooling via pragma style hints, however, HLS is not a silver bullet and whilst it has made the physical act of programming FPGAs much easier, one must still \emph{think dataflow} to obtain good performance \cite{ma2017optimizing}.

There have been a number of previous activities investigating the role that FPGAs can play in accelerating HPC codes, such as \cite{brown2019s} and \cite{lfricfpga}, and it is fair to say that whilst some kernels are suited for this type architecture, many struggle to perform well against the latest CPUs and GPUs. Nevertheless, it our belief that exploring the acceleration of popular HPC applications and mini-apps via reconfigurable architectures is an important activity. Not only does this aid in understanding the suitability of the technology for different codes and disseminate techniques for obtaining good performance on FPGAs, but furthermore it also enables hardware vendors to understand the mind-set of HPC software developers and their requirements.

\subsection{Vitis}

The Vitis Platform \cite{vitis} is an FPGA programming eco-system developed by Xilinx. First released in late 2019, this replaces Xilinx's SDAccel platform \cite{sdaccel} and promises to deliver an environment which lowers the barrier to entry in programming FPGAs and therefore their use in accelerating high performing codes, which is our interest here. 

For the work undertaken in this paper we use Vitis 2020.1, which at the time of writing is the latest version of the tool chain. As described above, the technology relies on HLS, where the programmer's C or C++ code is synthesised into the target RTL. Once the HLS code is written, the programmer then builds this via executing the \emph{v++} command in the terminal, which can be called by makefiles. Depending upon the flags provided, this command both synthesises HLS code and performs a linking stage. The later involves assembling the block design (shell) around the HLS kernel's IP block, and then executing Vivado which builds the bitstream in the normal manner.

Host side code uses OpenCL, which manages the transfer of data, launching of kernels, and marshalling of control. This is a convenient, standard, way of interacting with the FPGA and much more accessible than having to rely on vendor specific APIs. Furthermore, Vitis provides both software and cycle-accurate hardware emulation, which is driven by the provision of a single flag to the \emph{v++} tool and an environment variable. In this manner, it is convenient to develop and test functionality using emulation, relying predominantly on hardware runs and their long associated build time for performance measurements only. In addition to the building of code, Vitis also provides profiling support via Vitis Analyser, and a rich set of open source HLS libraries in the Xilinx Github repository. 

\subsection{Nekbone}
Nekbone \cite{ivanov2015evaluation} is a mini-app that captures the basic structure of the Nek5000 application \cite{nek5000-web-page}, which is a high order, incompressible Navier Stokes solver based on the spectral element method. A Gordon Bell prize winner, Nek5000 and its associated Nekbone mini-app are widely popular in the HPC community. Nekbone itself solves a standard Poisson equation using a Conjugate Gradient (CG) iterative method with a simple preconditioner on a block or linear geometry. This represents the principal computational kernel of Nek5000, and is a very useful tool for exploring the essential algorithmic elements that are pertinent to Nek5000, and many other HPC codes. Therefore, lessons learnt from the acceleration of Nekbone on FPGAs not only feed back to Nek5000, but are also of benefit to a wide variety of HPC codes that adopt a similar computational approach.

The solution phase in Nekbone consists of CG iterations, where each iteration involves vector operations, matrix-matrix multiply operations, nearest-neighbour communication, and MPI Allreduce operations. Of all of this functionality, the most expensive kernel is AX, which applies the Poisson operator and accounts for approximately 75\% of the overall Nekbone code runtime. It is this kernel that we focus on in this work, and all calculations are performed on an element by element basis, with each element consisting of a specific polynomial order configuration. For the runs conducted in this paper we use 800 elements, a polynomial order of 16 and three dimensions (this based upon a standard Nekbone test-case), which results in $16^3=4096$ grid points per element. In-fact, the AX kernel of Nekbone represents a challenging computational pattern, as each element requires a number of relatively small BLAS operations (in our configuration 16 x 16), resulting in, overall, very many small operations which tend to perform worse than a single, much larger BLAS operation \cite{dongarra2017design}. 

Listing \ref{lst:ax_kernel_cpu} illustrates a sketch of the code for this AX kernel, which is written in Fortran 77. The top level subroutine, \emph{ax}, accepts two integers, \emph{n} which is the polynomial order, and \emph{nelt} which is the number of elements. Four double precision floating point input arrays are also provided (the code is compiled with double precision, 64 bit, reals) and these are \emph{u}, \emph{g}, \emph{dxm1}, and \emph{dxtm1}. Lastly the \emph{w} argument is an output double precision floating point array which holds the results of the kernel. This subroutine loops over each element, explicitly calling the \emph{ax\_e} procedure for each at line 7. 

The \emph{ax\_e} subroutine itself is made up of three distinct parts, firstly a call to \emph{local\_grad3} at line 19, which performs matrix multiplications for the element in question, a local accumulations of values at lines 21 to 28, and finally a call to \emph{local\_grad3\_t} at line 30 which again performs matrix multiplications. Listing \ref{lst:ax_kernel_cpu} also illustrates the \emph{local\_grad3} procedure, where there are two matrix multiplications of $n^2$ by $n$ (lines 38 and 42), along with $n$ matrix multiplications of $n$ by $n$ at line 40. The \emph{local\_grad3\_t} procedure is omitted for brevity with the only difference being it contains two accumulations, before and after the final matrix multiplication. 

\begin{lstlisting}[frame=lines,caption={Sketch of AX kernel CPU code, applying the Poisson operator},label={lst:ax_kernel_cpu}, numbers=left]
subroutine ax(n, nelt, w, u, g, dxm1, dxtm1)
  integer, intent(in) :: n, nelt
  real(n,n,n,nelt), intent(in) :: u, g, dxm1, dxtm1
  real(n,n,n,nelt), intent(out) :: w
  
  do e=1, nelt
    ax_e(n, nelt, w(:,:,:,e), u(:,:,:,e), ...)
  enddo
end subroutine ax

subroutine ax_e(n, w, u, g, dxm1, dxtm1)
  integer, intent(in) :: n
  real(n,n,n), intent(in) :: u, g, dxm1, dxtm1
  real(n,n,n), intent(out) :: w
  
  real(n*n*n) :: ur, us, ut
  real :: wr, ws, wt
  
  call local_grad3(ur, us, ut, u, n, dxm1, dxtm1)
  
  do i=1,n*n*n
    wr = g(1,i)*ur(i) + g(2,i)*us(i) + g(3,i)*ut(i)
    ws = g(2,i)*ur(i) + g(4,i)*us(i) + g(5,i)*ut(i)
    wt = g(3,i)*ur(i) + g(5,i)*us(i) + g(6,i)*ut(i)
    ur(i) = wr
    us(i) = ws
    ut(i) = wt
  enddo
  
  call local_grad3_t(w, ur, us, ut, n, dxm1, dxtm1)
end subroutine ax_e

subroutine local_grad3(ur, us, ut, u, n, dxm1, dxm2)
  integer, intent(in) :: n
  real(n,n,n), intent(in) :: u, dxm1, dxm2
  real(n,n,n), intent(out) ::ur, us, ut
  
  call mxm(dxm1, n, u, n, ur, n*n)
  do k=0,n
    call mxm(u(:,:,k), n, dxtm1, n, us(:,:,k), n)
  enddo
  call mxm(u, n*n, dxtm1, n, ut, n)
end subroutine local_grad3

\end{lstlisting}


For the configuration we use in this paper (a polynomial order of 16), there are 831488 double precision floating point operations per element. Furthermore, different parts of the kernel are bound by different limits, with some aspects memory bound, and others compute bound. For instance, it can be seen in Listing \ref{lst:ax_kernel_cpu} that the data ordering consumption of array \emph{u} in \emph{local\_grad3} varies significantly between each of the three matrix multiplications, and with only one of the \emph{mxm} routines consuming the data consecutively. Based upon profiling on an Intel Xeon Platinum Cascade Lake (8260M) CPU, we found that 35\% of L1, and 10\% of L2, cache reads missed for the \emph{ax} kernel. When scaling the number of cores, we found that with a weak scaling experiment over all 24 CPU cores, whilst the full memory bandwidth was being used, memory throughput only increased 12 times compared to a single core.

\subsection{Hardware setup}

For the runs contained in this paper we use an Alveo U280 card, which contains an FPGA chip with 1.08 million LUTs, 4.5MB of on-chip BRAM, 30MB of on-chip URAM, and 9024 DSP slices. This PCIe card also contains 8GB of High Bandwidth Memory (HBM) and 32GB of DRAM on the board. For the experiments contained in this paper we exclusively use the HBM as our external memory store. All code is compiling at optimisation level three for both the host and device code, and GCC version 7.4. More details around the hardware setup can be found in the Artefact Description (Appendix \ref{sec:ad}).

\section{HLS kernel optimisation}
\label{sec:kernel_optimisation}
Our first step was to convert the Fortran 77 \emph{ax} kernel into C++, and apply the appropriate HLS pragmas to decorate arguments and set the AXI4 protocol on ports appropriately. We are maintaining double precision floating point in this work, and the code was synthesised by HLS, with the tooling generated correct target code for the FPGA, where results from runs on the Alveo U280 matched those on the CPU within error limits. Table \ref{fig-kernel-optimisation} illustrates the performance for different versions of this kernel, executed with a polynomial order of 16, and 800 elements. All results are averaged over three runs on the Alveo U280, and the \emph{Initial FPGA version} represents the performance obtained from this first attempt. The top line of Table \ref{fig-kernel-optimisation}, \emph{24 core Xeon Platinum CPU} is the performance of the kernel running over 24 cores of an Intel Xeon Platinum Cascade Lake (8260M). This is included to provide a comparison of FPGA performance, and at 0.03\% of the CPU performance, it can be seen that our initial FPGA kernel was very significantly slower than when running the existing, parallelised Nekbone, over all cores of the CPU.

For all FPGA versions we also include a \emph{\% theoretical performance} entry in Table \ref{fig-kernel-optimisation}. This is the theoretical best performance of the HLS algorithm at a specific clock frequency, assuming that the pipeline is fully filled. There are 203 double precision floating point operations required for each grid point of each element, and up to and including the \emph{optimise memory access} version of our kernel in Table \ref{fig-kernel-optimisation}, not all operations could run concurrently. Therefore, the percentages in the first three rows are calculated against a theoretical performance of only 6.9 GFLOPS. Starting at our fourth version, \emph{optimise matrix multiplications}, all operations could run concurrently and-so the theoretical performance increased 61 GFLOPS at 300MHz, which is the figure that the next three rows calculate against. At 400 Mhz this theoretical performance increased to 81.2 GFLOPS, which is the value used by the last row.

\begin{table*}[h]

 \centering
\begin{tabular}{ | c c c c | }
\hline
\textbf{Description} & \textbf{Performance (GFLOPS)} & \textbf{\% performance of CPU} & \textbf{\% theoretical performance} \\\hline
24 core Xeon Platinum CPU & 65.74 & - & - \\ \hhline {|=|=|=|=|}
Initial FGPA version & 0.020 & 0.03\% & 0.29\% \\ \hline
Optimised for dataflow & 0.28 & 0.43\% & 4.06\%\\ \hline
Optimised memory access & 0.42 & 0.63\% & 6.09\% \\ \hline
Optimise matrix multiplications & 12.72 & 19.35\% & 20.85\% \\ \hline
Ping-Pong buffering & 27.78 & 42.26\% & 45.54\% \\ \hline
Remove pipeline stalls & 59.14 & 89.96\% & 96.95\% \\ \hline
Increase to 400 Mhz & 77.73 & 118\% & 95.73\% \\ \hline
\end{tabular}
\caption{Performance of the FPGA AX kernel as different HLS optimisations were applied, for a polynomial order of 16 and 800 elements. Comparison to performance obtained by 24 cores of a Xeon Platinum (Cascade Lake) 8260M CPU is included for reference}
\label{fig-kernel-optimisation}
\end{table*}

The first version of our kernel was still very much based on the CPU code, and this illustrates that, whilst the HLS tooling is mature enough to accept C++ CPU-based code and synthesise this into something that will execute correctly, it will seldom provide good performance. Via the analysis pane, Vitis HLS provides detailed feedback around the potential performance of HLS codes, and we could see that the Initiation Interval (II), which is the number of cycles between each iteration entering a pipelined loop, was rather large. Ideally this number would be one, where every cycle a new iteration starts to progress, and in such a case once the pipeline is filled then an iteration completes every cycle. However, due to spatial dependencies this was 102 in the matrix multiplications, and therefore a significant performance limitation. We refactored the code, not only modifying the matrix multiplication algorithm itself, but also splitting the code up into separate HLS dataflow regions that could run concurrently. 

\begin{figure}[h]
\centering
\includegraphics[scale=0.40]{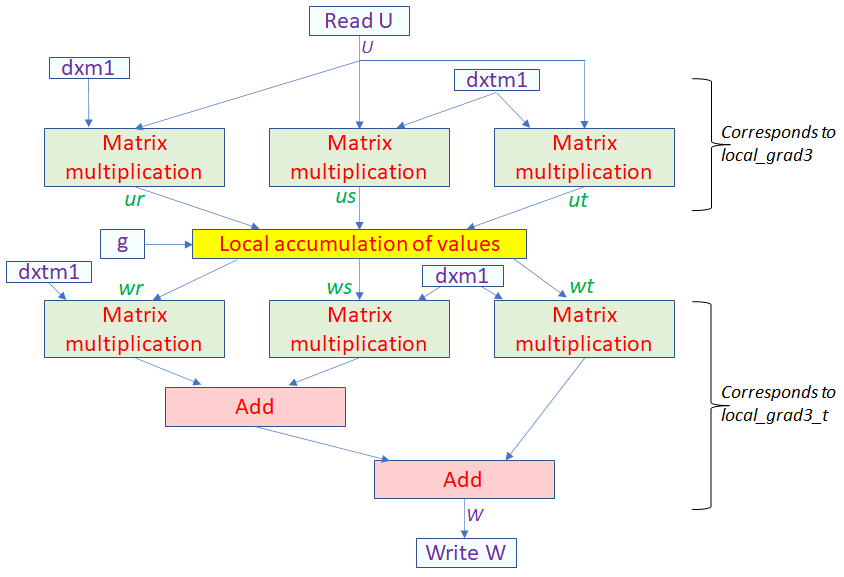}
\caption{Illustration of the dataflow architecture of our HLS kernel}
\label{fig:dataflow_architecture}
\end{figure}

Figure \ref{fig:dataflow_architecture} illustrates this new dataflow architecture, where each box corresponds to an HLS dataflow region, which are connected together via HLS streams (configured as FIFOs of depth 16) and all running concurrently. External data is read from HBM via an explicit dataflow region, for instance \emph{Read U} in Figure \ref{fig:dataflow_architecture} is a dataflow stage that reads in \emph{u} and then streams this to the three connected matrix multiplication dataflow regions. These, combined with the dataflow regions for reading in \emph{dxm1} and \emph{dxtm1} correspond to \emph{local\_grad3} in Listing \ref{lst:ax_kernel_cpu}. Result data from these first three matrix multiplications, \emph{ur}, \emph{us}, and \emph{ut}, is streamed to the local accumulation dataflow region (corresponding to lines 21-28 of Listing \ref{lst:ax_kernel_cpu}), and results from this step are streamed to the next three matrix multiplications which then feed their results into subsequent addition stages. Lastly, the result for each grid cell (with a polynomial order of 16 there are 4096 grid cells per element) is fed into the \emph{write w} stage, which writes the result back to HBM.

In addition to each HLS dataflow region of Figure \ref{fig:dataflow_architecture} running concurrently, reading in data via streams and streaming result data to the next stage, the matrix multiplication algorithm itself was refactored. This was modified to follow the style adopted by Xilinx in their open source Vitis BLAS library \cite{vitis-libraries}. Compared with the initial version, this work improved performance by over 10 times (\emph{optimised for dataflow} in Table \ref{fig-kernel-optimisation}), but whilst that in itself sounds impressive, the kernel was still over 200 times slower than the CPU version and achieving less than 5\% of the theoretical performance. 

Vitis Analyser provides profiling capabilities, and one metric reported was how efficiently the global memory and kernel(s) are interacting. From this profiling information it could be observed that we were achieving an aggregate bandwidth of only 952 MB/s (the HBM specification quotes 460 GB/s as the maximum bandwidth) and as such a tiny percentage of utilisation across our kernel ports. 

There were two reasons for this poor performance, firstly that kernel arguments shared one single port to one bank of the HBM. In-fact the HBM on the Alveo U280 is split up into 32 banks, each of 256MB, and fronted by 16 memory controllers, each with channels connected to two banks. Therefore, to obtain optimal performance one should utilise the entire HBM, rather than just one bank which is the default \cite{u280}. To address this we configured each kernel argument as a separate port and connected these to different HBM banks that were at-least a distance of two away from any others (to utilise separate memory controllers too). The second reason for the poor memory bandwidth was that the data-width of our ports was 64 bit (double precision floating point), and by increasing this to 512-bit, effectively fetching 8 double precision values per access, we drastically reduced the number of memory accesses required. The result was that Vitis Analyser now reported an average bandwidth of 95\% for memory accesses, which increased the performance of our kernel by approximately a third (\emph{optimised memory access} in Table \ref{fig-kernel-optimisation}).

However, at this point we were only achieving 0.6\% of the performance of the CPU, and so clearly there were still significant bottlenecks limiting performance. A sketch of our matrix multiplication code, based upon Xilinx's Vitis BLAS library, is illustrated in Listing \ref{lst:slow_mm}. However it can be seen that this does not start streaming out result values at line 11, until the last iteration of the outer loop \emph{k}. Therefore subsequent stages, such as the local accumulation and subsequent matrix multiplications are sitting idle until result data is streamed, which causes excessive wait times for each element.

\begin{lstlisting}[frame=lines,caption={Sketch of the matrix multiplication in our HLS kernel, which was based on the implementation in Xilinx's BLAS Vitis Library},label={lst:slow_mm}, numbers=left]
double a_temp[NX], c_temp[NX*NX][NX];
#pragma HLS array_partition variable=a_temp complete

for (int k = 0; k < NX; k++) {
  for (int j=0;j<NX*NX;j++) {
    double b_val=b.read();
    for (int i=0;i<NX;i++) {
      if (k==0) c_temp[j][i]=0.0;
      if (j==0) a_temp[i]=a.read();
      c_temp[j][i]+=a_temp[i] * b_val;
      if (k==NX-1) c.write(c_temp[j][i]);
    }
  }
}
\end{lstlisting}

Fundamentally, the issue was that whilst we had split our code into HLS dataflow regions, algorithmic issues where still limiting what parts could run concurrently. Listing \ref{lst:fast_mm} illustrates a refactored version, where we brought the outer loop inside the inner, \emph{i}, loop and manually unrolled it. We have to load all \emph{b} values for the inner loop, which are the statements at lines 5 to 9, between the two for loops (the \emph{packaged\_double} structure packs eight double precision reals, therefore lines 6 and 8 are loading 8 values with a single read which are then unpacked into the \emph{b\_temp} array.) This new version provides two advantages, firstly there is no longer the delay in streaming results, as once the pipeline is filled then a result is generated every cycle, significantly increasing the occupancy of subsequent stages. Secondly, as we have manually unrolled the loop over \emph{k}, there are also many more floating point operations running concurrently. Compared to Listing \ref{lst:slow_mm}, which operates on one floating point operation per cycle, the algorithm in Listing \ref{lst:fast_mm} performs 31 floating point operations per cycle. Bearing in mind that there are six matrix multiplication dataflow regions, this concurrency of arithmetic operations is significant and resulted in a performance improvement of over 30 times, boosting performance to 12.72 GFLOPS (\emph{optimise matrix multiplications} in Table \ref{fig-kernel-optimisation}). Due to this additional vectorisation, where it was now possible to run all 203 floating point operations concurrently for the first time, the kernel's theoretical performance also increased, from 6.9 to 61 GFLOPS.

\begin{lstlisting}[frame=lines,caption={Sketch of the refactored matrix multiplication, which generates results every cycle once filled},label={lst:fast_mm}, numbers=left]
double a_temp[NX][NX], b_temp[NX];
#pragma HLS array_partition variable=a_temp dim=1 complete

for (int j=0;j<NX*NX;j++) {
  // Load b values that are needed by the inner loop
  struct packaged_double in_data=b.read();
  for (int q=0;q<8;q++) b_temp[q]=in_data.data[q];
  in_data=b.read();
  for (int q=0;q<8;q++) b_temp[q+8]=in_data.data[q];
  
  for (int i=0;i<NX;i++) {
    if (j==0) a_temp[0][i]=a[0].read();
    double temp_0=a_temp[0][i] * b_temp[0];

    if (j==0) a_temp[1][i]=a[1].read();
    double temp_1=a_temp[1][i] * b_temp[1];

    ....
    c.write(C_temp_0 + C_temp_1);
  }
}
\end{lstlisting}

However, at this point the HLS kernel was still around five time slower than the AX kernel running over 24 cores of the CPU. To understand why we need to explore the HLS kernel in more depth, as it is slightly more complex than previously described. The groups of three matrix multiplications consume their input data (either \emph{u} for the first group, or \emph{wr}, \emph{ws}, and \emph{wt} for the second group) in different orders to each other. For performance, we read each grid point of \emph{u} only once, and do so contiguously to ensure that HLS imposes only one (expensive) read request for each element. However, different matrix multiplications require consumption of this data in a different order. The same issue exists for the second group of matrix multiplications, where these kernels require their input data in a different order than \emph{wr}, \emph{ws}, and \emph{wt} are generated. As such, each matrix multiplication kernel is associated with a buffer which holds the all grid points for an element (in our configuration 4096 double precision floating point values). These are filled up and, once full, data is then served from the buffers into their respective matrix multiplication kernels in the specific order required. 

Therefore, whilst the previous optimisation step had significantly increased the performance of our matrix multiplication kernel itself, due to this buffering, there were still overheads. Effectively, the code was working implicitly in three phases for each element. The first phase was where data was read into the buffers of the first matrix multiplications, with no calculations active, which for our problem size resulted in 512 cycles of inactivity whilst the buffers were filled (due to the 512 bit width of data ports, we filled eight values per cycle). In the second phase the first bank of matrix multiplications and the local accumulation were active for the element, with intermediate result data streaming into the buffers of the last three matrix multiplications, and the third where the last three matrix multiplications and two addition dataflow regions were active. Therefore there was still a significant amount of wasted concurrency due to this buffering.

\begin{figure}[h]
\centering
\includegraphics[scale=0.40]{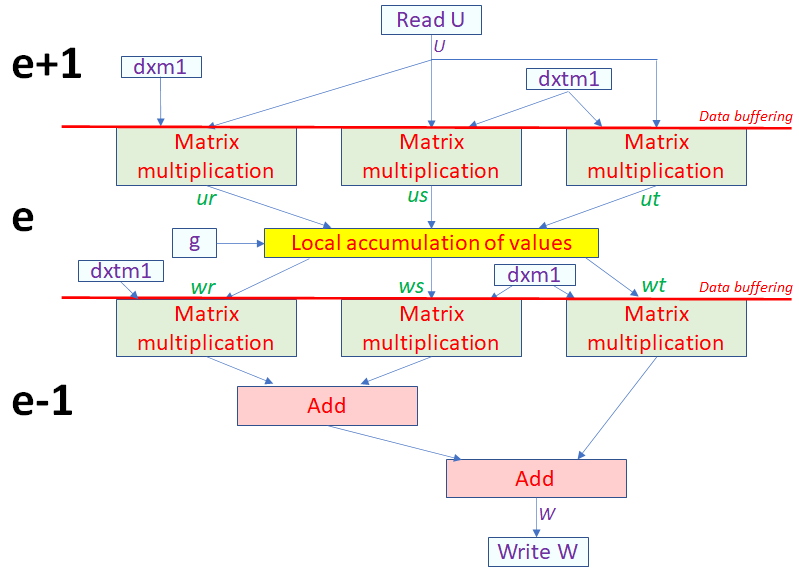}
\caption{Illustration of dataflow architecture running concurrently, by each of the three phases operating on different elements to avoid dependency issues}
\label{fig:dataflow_architecture_buffering}
\end{figure}

The challenge is that there is a dependency, where all data must be read in before the first group of matrix multiplications can operate for an element, and all intermediate results from the local accumulation must be present before the second group of matrix multiplication can run for the element. To address this we can run these three phases concurrently, but for different elements, which is illustrated in Figure \ref{fig:dataflow_architecture_buffering}. The first group of three matrix multiplications and local accumulation, is running for the current element, \emph{e}, the initial reading of data is occurring for the next element, \emph{e+1}, and the second group of three multiplications and two additions for the previous element, \emph{e-1}. In this manner, we can ensure that all three phases are running concurrently, thus making more effective use of the hardware.

\begin{figure}[h]
\centering
\includegraphics[scale=0.40]{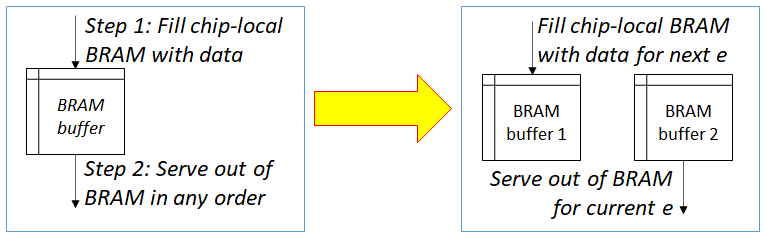}
\caption{Illustration of change to the data buffering and reordering code}
\label{fig:double_buffering}
\end{figure}

Figure \ref{fig:double_buffering} illustrates the change that was required to the buffering itself in order to support this concurrency. Previously the buffering had worked in two steps, first reading in the data contiguously (512 cycles for a polynomial order of 16) and then once completed, serving the data out in the order required by the specific matrix multiplication. The code was modified to follow what software developers term a \emph{double buffering approach}, where there are two buffers working cyclically. The first buffer is filled up with data for the next element, whilst the second buffer concurrently serves data for the current element. Once this is completed, the buffers swap and the process continues for the next element, hence data can be both read in contiguously, and delivered out of order, concurrently. 

An important aspect to highlight is that, coming from a software development rather than hardware design background, there was some tension here initially. We first developed this double buffering approach manually in HLS, which resulted in significant resource usage. Fundamentally, we did not realise that the use of ping-pong buffers between dataflow regions in HLS is effectively the same as what we were trying to implement manually via double buffering. By adopting HLS's native approach with two separate dataflow regions connected via PIPO buffers, rather than our own, we obtained a much more effective implementation, both in terms of resource usage and performance. 

The impact of this optimisation is illustrated in Table \ref{fig-kernel-optimisation} by the row \emph{ping-pong buffering}, which more than doubled performance to 27.78 GFLOPS. However, we were still achieving less than half the performance of the CPU and just less than half of our kernel's theoretical performance. Via the analysis pane of Vitis HLS, we discovered that there was a further issue around how HLS was synthesising the matrix multiplication algorithm. The way in which the code of Listing \ref{lst:fast_mm} was structured meant that there was a dependency between loading the \emph{b} values into \emph{b\_temp} at lines 5 to 9, and then consuming them in the inner loop. This caused HLS to drain the pipeline of the inner loop for every iteration of \emph{j}. The pipeline depth was 45, and the inner loop was only iterating to 16, hence this continued draining and filling was very expensive, and at no point was the pipeline of the inner loop fully filled. 

We modified this problematic code by moving the reading of \emph{b} into the inner loop and on an element by element basis. By doing so, this removed the inter-loop dependency, and HLS was able to keep the pipeline of the inner loop filled. For our problem size this meant going from 204800 batches of 16 cycles (having to drain between each batch), to 1 batch of 3276800 cycles, with the pipeline filled for over 99.9\% of the time. This provided a further significant increase in performance to 59 GFLOPS.

At this point we were achieving around 90\% the performance of the 24 core Xeon Platinum CPU. One should consider that the theoretical performance of our HLS kernel was 61 GFLOPS, of which we were achieving almost 97\%. Therefore, based upon the current configuration, it was never going to be possible for our HLS kernel to match or exceed the CPU's performance. As such, we had to explore options which would increase the kernel's theoretical performance itself.

On the Alveo U280 the default clock frequency is 300Mhz, which to change in Vitis involves only a configuration option. However, this should not be seen as a silver bullet because, without a well performing kernel in the first place, then a clock frequency increase might improve performance slightly, but won't address any underlying issues. Furthermore, an increased clock frequency impacts the overall complexity of the kernel, and by increasing to 400Mhz the depth of our matrix multiplication increased to 61 cycles. We found empirically that 400Mhz was the optimal clock frequency, and beyond this the complexity of the matrix multiplications increased very significantly, with the pipeline II increased to two. It was possible to reduce this back down to one by using the \emph{BIND\_OP} HLS pragma to increase the latency of the double precision floating point cores, but the performance we obtained by doing so never matched that of 400Mhz.

By increasing the clock frequency to 400Mhz, we achieved performance of over 77 GFLOPS, which is faster than Nekbone running on the 24-core CPU. This also achieved around 96\% of the theoretical performance of the kernel, which demonstrates that once filled, the dataflow stages and pipelines within, are staying well utilised. Therefore, with data continually feeding into our HLS kernel and results streaming out, by moving to a dataflow approach we have an algorithm which is fully utilising the hardware and no longer memory bound. 

\section{Multiple kernel performance}
\label{sec:multiple_kernel}
In Section \ref{sec:kernel_optimisation} we explored the optimisation of our AX HLS kernel, ultimately resulting in a code running at 400MHz that outperforms a 24 core Cascade Lake Xeon Platinum CPU by around 18\%. However, our focus thus far has been on optimising a single kernel on the U280, and this only utilises a fraction of the overall FPGA's resources. Therefore, there was further opportunity to increase performance by leveraging multiple kernels and as each element is independent from any other element, these can be distributed across the kernels.

The resource usage of our initial single kernel version was rather unbalanced. At the Super Logic Region (SLR) level Vitis HLS estimated it required 115\% of the SLR's on-chip BRAM, but only 34 \% of the DSP slices, 24 \% of the Flip Flops, 28\% of the LUTs\ and none of the on-chip UltraRAM (URAM). The U280 has three SLRs, but as we aimed to scale up to multiple kernels then clearly the amount of BRAM consumed would be a major limitation. 

Therefore to address this we used the \emph{BIND\_STORAGE} HLS pragma of Vitis HLS (which replaces the \emph{RESOURCE} pragma of Vivado HLS) to direct what hardware components should be used to implement which areas of memory. We allocated the FIFO queues associated with the HLS streams, and arrays associated with the data-reordering ping pong buffers into LUTRAM, and placed the data storage associated with each matrix multiplication dataflow region into the on-chip URAM. This resulted in a more balanced resource usage, with the number of DSP slices unchanged, the SLR BRAM usage reduced down to 32\%, the FF usage increased to 29 \%, LUT usage to 36\%, and URAM to 30\%. We observed a negligible impact in performance by making this change, but crucially were now in a position to scale up the number of kernels.

\begin{figure}[h]
\centering
\includegraphics[scale=0.40]{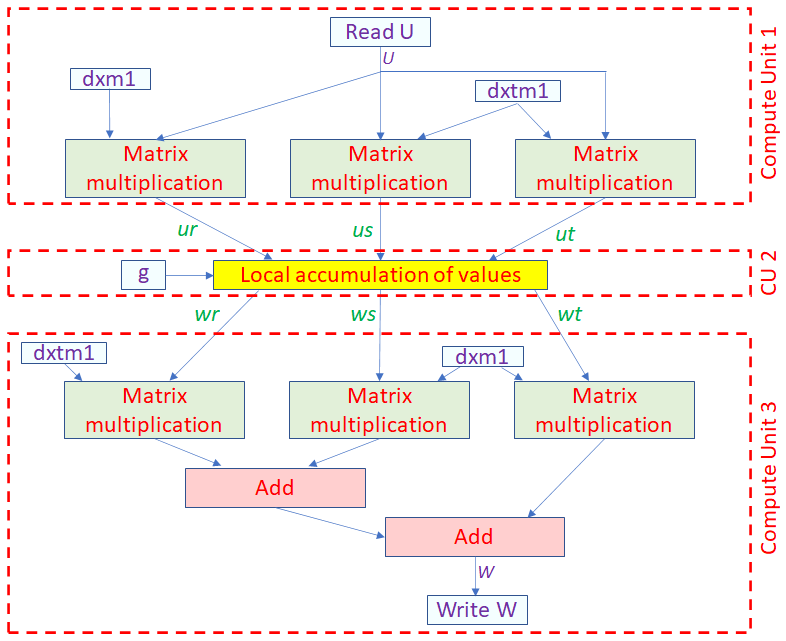}
\caption{Illustration of splitting apart of kernel into three separate Compute Units (CUs)}
\label{fig:kernel_three_cu}
\end{figure}

However there was a further issue because, when we increased the number of HLS kernels, also known as Compute Units (CUs), to two, initial performance was very poor. From studying the logs of Vivado implementation, it was observed that the tooling was dynamically down clocking the kernels from 400MHz to less than 200MHz in order to meet timing. After some experimentation it was found that this could be fixed by splitting each HLS kernel up into three separate CUs, connected by AXI4 streams. This is illustrated in Figure \ref{fig:kernel_three_cu}, where the first CU contains the first three matrix multiplications of Figure \ref{fig:dataflow_architecture}, the second CU performs the local accumulation of values, and the third CU contains the last three matrix multiplications and two additions. 

This updated version initially resulted in routing errors due to congestion in the matrix multiplication of the first CU. We had not encountered this before, and whilst instructing Vitis to use the \emph{Congestion\_SpreadLogic\_high} implementation strategy fixed the issue, it too resulted in poor performance. After some investigation it was found that there were naming conflicts between the first and third CUs. Specifically, that the names of the matrix multiplication functions were the same in each CU, and clearly the place and route step was attempting to perform some optimisation by consolidating these already complex components together, which resulted in significant congestion. The fix was to uniquely name the functions. The bit-stream generation time (synthesis, place, and route) depended heavily on the number of kernels. One kernel took around six hours to build, however this increased sharply to around thirty hours for four kernels.


Table \ref{fig-overall-performance} contains a performance and power efficiency comparison of multiple kernels (each comprised of the three CUs) against other technologies. To obtain the power draw figures, Xilinx's \emph{xbutil} was used for the FPGA which returns the total power being consumed by the card, Intel's 
Running Average Power Limit (RAPL) for the CPU, and \emph{nvidia-smi} for the GPU. The CPU is the same 24 core Xeon Platinum Cascade Lake (8260M) compared against in Section \ref{sec:kernel_optimisation}, and running over all 24 cores resulted in an power efficiency of 0.37 GFLOPS/Watt. For comparison we also include a single core CPU run, which resulted in 5.38 GFLOPS and power efficiency of 0.08 GFLOPS/Watt. 


\begin{table}[h]

 \centering
\begin{tabular}{ | c c c c | }
\hline
\textbf{Description} &  \makecell{\textbf{Performance} \\ \textbf{(GFLOPS)}} &  \makecell{\textbf{Power usage} \\ \textbf{(Watts)}} &  \makecell{\textbf{Power Efficiency} \\ \textbf{(GFLOPS/Watt)}} \\\hline
1 core of CPU & 5.38 & 65.16 & 0.08 \\
24 cores of CPU & 65.74 & 176.65 & 0.37 \\
V100 GPU & 407.62 & 173.63 & 2.34 \\
1 kernel & 74.29 & 45.61 & 1.63 \\
2 kernels & 146.94 & 52.47 & 2.80 \\
4 kernels & 289.02 & 71.98 & 4.02 \\
\hline
\end{tabular}
\caption{Performance and power comparison between different numbers of FPGA kernels on the Alveo U280, and runs on the CPU and GPU}
\label{fig-overall-performance}
\end{table}

Nekbone has mature support for GPU acceleration of the \emph{ax} kernel via CUDA \cite{gong2016nekbone}. We ran an experiment on a NVIDIA Tesla V100 GPU, compiling with the Portland Group Compiler version 20.5-0, and CUDA 10.2 . This resulted in 407 GFLOPS and, due to the high performance, a power efficiency of 2.34 GFLOPS/Watt. The GPU's performance is impressive, although it should be noted that the bespoke GPU acceleration in Nekbone has been developed and tuned over many years and GPU generations.

In Table \ref{fig-overall-performance} we report the performance for different numbers of our \emph{ax} kernels on the Alveo U280 FPGA. It should be noted that the performance for one kernel is slightly lower than that described in Section \ref{sec:kernel_optimisation}, which is due to the splitting apart of the kernel as discussed earlier in this section and the associated overhead of streaming data via AXI4 streams between CUs. One kernel draws 45.61 Watts (the FPGA idle with the bitstream loaded draws 39 Watts), and whilst the power efficiency of 1.63 GFLOPS/Watt of a single kernel is significantly higher than the CPU, it is somewhat disappointing when compared against the GPU. 

However, the advantages of FPGAs starts to become more apparent as we scale the number of kernels. We can fit up to four of our kernels on the U280, and at this configuration we achieve 289 GFLOPS. This over four times the performance of the 24 core CPU, and 71\% of the performance of the V100 GPU. The power consumption of four kernels is 72 Watts and it can be observed that, on average, adding an extra kernel requires approximately an additional 7 Watts, with a performance increase close to 74 GFLOPS per kernel. With four kernels, the power efficiency is over 4 GFLOPS/Watt, which is significantly higher than that of the GPU.

We were pleasantly surprised with how well FPGA performance scaled as we added kernels. In part this is because the ports of each kernel connect to different HBM banks, so there is no contention. We found that if HBM banks were shared between kernels then it resulted in hold conflicts, excessive time in the additional hold fix phase of routing (over 12 hours) and reduced performance. Effectively, by keeping the kernels separate they can then run independently and scale better.

\section{Conclusions and further work}
\label{sec:conclusions}

In this paper we have explored the porting of Nekbone's AX kernel to FPGAs using Vitis. Nekbone is a popular mini-app in the HPC community, and we have described the steps required to transform the algorithm which accounts for over 75\% of the runtime, from a CPU-based Von Neumann style to dataflow form. It is stark that on the FPGA, the performance difference by doing so is approximately 4000 times. Furthermore, we found that calculating theoretical performance was a useful metric as it enabled us to gauge how much more opportunity there was to optimise the kernel, and provide additional context to the comparison against the CPU performance.

We compared the performance and power efficiency of up to four of our HLS kernels on an Alveo U280 against that of a 24 core Intel Xeon Platinum Cascade Lake CPU, and a NVIDIA Tesla V100 GPU. In comparison to the CPU, four HLS kernels provides over four times the performance, at around two and a half times less power consumption. By contrast, the GPU is more challenging to match, and the performance of Nekbone on the V100 GPU is impressive, which is a result of both the many years of effort developing the Nekbone GPU kernel, and furthermore significant effort expended by NVIDIA in their tooling and V100 hardware. Therefore, we consider the results obtained optimistic for the future of FPGAs in HPC, as not only is the power efficiency of the FPGA version almost double what the GPU can provide, but given a slightly larger FPGA that could fit an extra kernel then the performance gap would be much narrower, and two extra FPGA kernels would outperform the GPU.

In terms of further work, exploring the role of reduced precision would be interesting. All reals are double precision thus far, but by moving to single, half or fixed point, then this will not only reduce the number of DSP slices required for matrix multiplication calculations, but will also reduce the memory and potentially LUT requirements too. By doing so it is likely that we will be able to increase the number of kernels, and it will be very interesting to explore how this either closes or widens the performance gap against GPUs (which themselves are also optimised for reduced precision.) Furthermore, we have not focused on the overhead of data transfer to or from the FPGA, or GPU, card in this paper. This is because one data transfer tends to occur for around 200 invocations of the kernel, so the cost of the transfer is negligible and ameliorated by the repeated executions on the accelerator. However, communication is required between the invocations of the kernel when run on multiple nodes. In this paper we focused on the kernel itself, which does not involve communication, and on a single node. But it would be very interesting to extend and explore running across multiple nodes, and the FPGA to FPGA communication that would be required between invocations, especially compared against GPU to GPU communications. 

We conclude that there are some very exciting developments in the FPGA community at the moment. Even a few years ago, the ability to develop an HLS kernel and scale it up, that could convincingly outperform a modern Xeon CPU would be unrealistic. We believe that with a slightly larger FPGA, such as the future Versal architecture, then performance will be more competitive against next generation GPUs and the already demonstrable power efficiency advantages of current generation technologies are noteworthy. Given the trajectory of FPGA vendors, such improvements in the coming years is a realistic proposition, and this demonstrates that in the medium term reconfigurable architectures could become much more mainstream in next generation supercomputers.

\section{Acknowledgements}
The authors would like to thank Xilinx for the donation of the Alveo U280 card used throughout the experiments of work. This work was funded under the EU EXCELLERAT CoE, grant agreement number 823691.


\bibliographystyle{./bibliography/IEEEtran}
\bibliography{./bibliography/IEEEabrv,./bibliography/IEEEexample}

\appendices

\section{Artifact Description Appendix}
\label{sec:ad}

\subsection{Description}

\subsubsection{Check-list (artifact meta information)}

{\small
\begin{itemize}  
  \item {\bf Program: } C++ and Fortran
  \item {\bf Compilation: }GCC version 7.4 with -O3, Vitis version 2020.1. On the GPU we used the Portland Group Compiler version  20.5-0, and  CUDA  10.2.
  \item {\bf Data set: }Runs were based on the \emph{nek\_gpu1} testcase configuration in the GPU branch of the Nekbone Github repository. As described in the paper, the number of elements was increased to 800.
  \item {\bf Run-time environment: } A variety of machines were used for comparison, all running Linux. For the Alveo U280 the latest environment at the time of writing was used (XRT 202010.2.6.655, xdma and xdma-dev 201920.3)
  \item {\bf Hardware: } We used a Xilinx Alveo U280 for the FPGA runs, this is hosted by a system with a Xeon Platinum Skylake (8170) and 192GB RAM. For CPU comparison runs we ran on a Xeon Platinum Cascade Lake (8260M) processor with 192GB RAM. For GPU runs we ran on the Cirrus tier-2 UK HPC machine, which provides a NVIDIA Tesla V100-SXM2-16GB (Volta) GPU, hosted by two Intel Xeon Gold Cascade Lake (6248) CPUs and 384 GB RAM.
  \item {\bf Binary: } Nekbone CPU versions require MPI. Xilinx Vitis library is required to synthesise the kernel and generate the bitstream.
  \item {\bf Execution: } We built and executed all executables on Linux.
  \item {\bf Output: } Nekbone provides a performance measure in GFLOPS and we also added in additional timing and manually calculated to ensure that the reported value was correct.
  \item {\bf Publicly available?: }Not yet
\end{itemize}
}

\subsubsection{Hardware dependencies}
Any machine running Linux with appropriate Alveo U280 FPGA PCIe card installed
\subsubsection{Software dependencies}
The latest version of Nekbone from the Github repository, GCC version 7.4, the support libraries installed for the board and the Vitis platform. 
\subsubsection{Datasets}
Runs were based on the \emph{nek\_gpu1} testcase configuration in the GPU branch of the Nekbone Github repository.
\subsection{Installation}
We synthesised our kernel using Vitis HLS via the \emph{v++} command. It is also possible to synthesise directly via the HLS IDE, and this was useful for steps which involved leveraging the analysis pane. The \emph{v++} command was then used to link, which assembles the shell and calls out to Vivido to generate the bitstream. The host code was written in OpenCL (the appropriate libraries ship with Vitis) and launching our bitstream simply involved executing the host code, which via the appropriate OpenCL calls programmed the device as appropriate. 

\subsection{Experiment workflow}
\begin{enumerate}
\item Develop the appropriate HLS kernel 
\item Use Vitis HLS to synthesise this and generating corresponding \emph{.xo} files
\item Use Vitis HLS in linking mode to generate the bitstream \emph{.xclbin} file
\item Compile the host OpenCL code using GCC
\item Execute the host code, which will launch the bitstream
\item Optionally, enable profiling and after the run use Vitis Analyser to explore this information.
\end{enumerate}

\subsection{Evaluation and expected result}
We compared our results against the CPU and GPU version of Nekbone. On the CPU this ran across all 24 cores of the 8260M, and whilst we did develop an OpenMP version, we in-fact found that the existing MPI code with an MPI process per core provided best performance (there is no communication inside the AX kernel itself) and-so this is the code that was used. For the GPU version, Nekbone provides both an OpenACC and CUDA implementation of the AX kernel. We found the performance difference between these two fairly negligible, but the CUDA code was marginally faster and hence this is the one used in this paper. For our HLS code, we first converted Nekbone's Fortran 77 into C++. However, we found that on the CPU and GPU, the Fortan 77 gave better performance, and hence it is this original version of the code that was used for performance comparisons. All results have been checked at the grid point level to ensure that they are producing consistent results between the different versions of the code. All results reported in this paper are averaged over three runs.

\subsection{Experiment customization}
It is, of course, possible to experiment with the kernels and use these to run different system sizes, for instance modifying the polynomial order or number of elements. On an Alveo U280 the maximum number of kernels we could fit was four, and with a larger FPGA such as the future Versal architecture, then this could be scaled up further.

\section{Artifact Evaluation}

\subsection{Results Analysis Discussion}
In the host code we use OpenCL's profiling capability which provides microsecond resolution timings for event (kernel execution) starting and ending. We also implemented manual timing via the \emph{gettimeofday} call, to provide a second timing comparison point and ensure what OpenCL reported was correct (both approaches to timing matched very closely.) All results were checked, grid point by grid point, for consistency between the FPGA and CPU versions to ensure that they are calculating the same quantities and we were undertaking a fair experiment. For all experiments runtimes were averaged over at-least three runs, and power consumption figures were reported by XRT for the FPGA, RAPL for the CPU, and \emph{nvidia-smi} on the GPU.

\end{document}